\documentclass[prb,twocolumn,superscriptaddress,a4paper,twoside]{revtex4}

\usepackage{amsmath}
\usepackage{amssymb}
\usepackage{todonotes}
\usepackage{nicefrac}
\usepackage{graphics}
\usepackage{color}
\usepackage{graphicx}
\UseRawInputEncoding

\makeatletter
\newcommand*{\rom}
[1]{\expandafter\@slowromancap\romannumeral #1@}
\makeatother

\begin{document}

\title{Spectral functions of the half-filled 1D Hubbard chain within
the exchange-correlation potential formalism}

\author{F. Aryasetiawan and T. Sj\"ostrand}
\affiliation{Department of Physics, Division of Mathematical Physics, 
Lund University, Professorsgatan 1, 223 63, Lund, Sweden}

\begin{abstract}
The spectral functions of the one-band half-filled 1D Hubbard chain are calculated using the 
exchange-correlation potential formalism developed recently. 
The exchange-correlation potential is adopted from the exact
potential derived from the Hubbard dimer. Within an approximation in which the full 
Green function is replaced by a non-interacting one, the spectral functions can be calculated 
analytically. Despite the simplicity of the approximation, the resulting spectra are in favorable 
agreement with the more accurate results obtained from the dynamic density-matrix renormalization 
group method. In particular, the calculated band gap as a function of $U$ is in close agreement 
with the exact gap obtained from the Bethe ansatz. In addition, the formal general 
solution to the equation of motion of the Green function is presented and the difference between 
the traditional self-energy approach and the exchange-correlation potential formalism is also 
discussed and elaborated. A simplified Holstein Hamiltonian is considered to further illustrate 
the general form of the exchange-correlation potential. 
\end{abstract}

\maketitle

\section{Introduction}

Very recently, a completely different route for calculating the Green function
was proposed. The formalism replaces the traditional self-energy by a
time-dependent exchange-correlation potential, 
$V_{\mathrm{xc}}(r,r^{\prime};t)$, 
which acts as a multiplicative potential on the Green function, in
contrast to the self-energy which acts as a convolution in space and time 
\cite{aryasetiawan2022}.

In this previous work, the formalism was illustrated by deriving the exact
$V_{\mathrm{xc}}$ of the Hubbard dimer expressed in the site orbitals.
In this paper, the formalism is applied to calculate the Green function of the
one-dimensional Hubbard chain at half filling by utilizing the $V_{\mathrm{xc}}$
derived from the Hubbard dimer, rewritten in the bonding and anti-bonding
orbitals appropriate for extension to the one-dimensional chain. 
Under certain approximations, the Green function
can be calculated analytically, providing explicit insights into the dependence
of the spectra on the parameters of the model. Despite the simple appearance of
the Hamiltonian of the 
one-dimensional Hubbard chain, it is actually a rather stringent test for 
a many-electron theory. It is well known that the exact solution based on the Bethe ansatz
yields two branches of collective excitations corresponding to the holon and the spinon 
\cite{giamarchi2004,essler2005}.
While the Bethe-ansatz solution provides the dispersion of the holon and the spinon, it does
not contain information about the spectral distribution. The spectral
functions, however, have been calculated within the dynamical density-matrix renormalization
group method \cite{benthien2007} and furnish comparison with the present work.
Earlier calculations are limited to the large $U$ limit \cite{sorella1992,penc1996} and
there are also related investigations on Luttinger liquid \cite{meden1992,voit1993}.
The 1D Hubbard model is physically relevant as shown, for example, in the case of SrCuO$_2$, which
has been found to exhibit the spinon and holon excitations \cite{kim1996,kim1997}.

Apart from application to the 1D Hubbard chain, a
formal solution of the equation of motion of the Green function
is derived, which provides an iterative scheme
in practical calculations. 

To contrast the traditional self-energy approach and the exchange-correlation formalism,
comparison between the two is made and discussed from a historical perspective
and the differences are emphasized and elaborated. 

In addition, the exchange-correlation potential of a simplified Holstein Hamiltonian, 
describing a coupling between a core electron 
and a set of bosons (e.g., plasmons or phonons) is derived analytically 
and serves as a further illustration of the exchange-correlation potential formalism.
This sheds
lights on the form of $V_{\mathrm{xc}}$, which offers a very simple physical 
interpretation. 

\section{Theory}

The exchange-correlation potential formalism developed in the previous work
\cite{aryasetiawan2022} is summarized and the main results are presented.

\subsection{Equation of motion of the Green function}

The zero-temperature time-ordered Green function is defined as
\cite{fetter}%
\begin{equation}
iG(rt,r^{\prime}t^{\prime})=\left\langle T[\hat{\psi}(rt)\hat{\psi
}^{\dag}(r^{\prime}t^{\prime})]\right\rangle ,
\end{equation}
where $r=(\mathbf{r},\sigma)$ labels both space and spin variables, $\hat
{\psi}(rt)$ is the Heisenberg field operator, $T$ is the
time-ordering symbol, and $\left\langle ...\right\rangle $ denotes expectation
value in the ground state. Extension to finite temperature is quite
straightforward. The many-electron Hamiltonian defining the Heisenberg
operator is given by%
\begin{align}
\hat{H}  &  =\int dr \;\hat{\psi}^{\dag}(r)h_{0}(r)\hat{\psi
}(r)\nonumber\\
&  +\frac{1}{2}\int drdr^{\prime}\;\hat{\psi}^{\dag}(r)\hat{\psi}^{\dag
}(r^{\prime})v(r-r^{\prime})\hat{\psi}(r^{\prime})\hat{\psi}(r),
\label{Hamiltonian}
\end{align}
where $h_{0}=-\frac{1}{2}\nabla^{2}+V_{\mathrm{ext}}(r)$ and $v(r-r^{\prime
})=1/|\mathbf{r-r}^{\prime}|$. In our notation, $\int dr=\sum_{\sigma}\int
d^{3}r$ and atomic units are used throughout, in which the Bohr radius $a_{0}%
$, the electron mass $m_{e}$, the electronic charge $e$, and $\hbar$ are set
to unity. For a system in equilibrium, the Hamiltonian is time independent and
$t^{\prime}$ may be set to zero. The equation of motion of the Green function
is given by%
\begin{align}
&  \left(  i\frac{\partial}{\partial t}-h_{0}(r)\right)
G(r,r^{\prime};t)\nonumber\\
&  +i\int dr^{\prime\prime}v(r-r^{\prime\prime})\left\langle T[\hat{\rho
}(r^{\prime\prime}t)\hat{\psi}(rt)\hat{\psi}^{\dag}(r^{\prime
})]\right\rangle \nonumber\\
&  =\delta(t)\delta(r-r^{\prime}). \label{EOM-G1}%
\end{align}
Since the Green function was conceived in the 1950's, the traditional approach
is to introduce the self-energy $\Sigma$ and truncate the
hierarchy of higher-order Green functions such that \cite{fetter}%
\begin{align}
&  -i\int dr^{\prime\prime}v(r-r^{\prime\prime})\left\langle T[\hat{\rho
}(r^{\prime\prime}t)\hat{\psi}(rt)\hat{\psi}^{\dag}(r^{\prime
})]\right\rangle \nonumber\\
&  =V_{\mathrm{H}}(r)G(r,r^{\prime};t)+\int dr^{\prime\prime}dt^{\prime\prime
}\Sigma(r,r^{\prime\prime};t-t^{\prime\prime})G(r^{\prime\prime},r^{\prime
};t^{\prime\prime}). \label{def:Sigma}%
\end{align}
The equation of motion then becomes%
\begin{align}
&  \left(  i\frac{\partial}{\partial t}-h(r)\right)  G(r,r^{\prime
};t)\nonumber\\
&  -\int dr^{\prime\prime}dt^{\prime\prime}\Sigma(r,r^{\prime\prime
};t-t^{\prime\prime})G(r^{\prime\prime},r^{\prime};t^{\prime\prime
})\nonumber\\
&  =\delta(r-r^{\prime})\delta(t), \label{EOM-Sigma}%
\end{align}
{where }$h=h_{0}+V_{\mathrm{H}}$ with $V_{\mathrm{H}}$ being the Hartree potential.

As shown in the previous work \cite{aryasetiawan2022}, it is possible to define an
exchange-correlation potential $V_{\mathrm{xc}}$ so that the Green function
fulfills the following equation of motion:
\begin{equation}
\left(  i\frac{\partial}{\partial t}-h(r)-V_{\mathrm{xc}}(r,r^{\prime
};t)\right)  G(r,r^{\prime};t)=\delta(r-r^{\prime})\delta(t), \label{EOM}%
\end{equation}
where $V_{\mathrm{xc}}$ is the Coulomb potential of the time-dependent
exchange-correlation hole, $\rho_{\mathrm{xc}}$:%
\begin{equation}
V_{\mathrm{xc}}(r,r^{\prime};t)=\int dr^{\prime\prime}v(r-r^{\prime\prime}%
)\rho_{\mathrm{xc}}(r,r^{\prime},r^{\prime\prime};t).
\end{equation}
and $\rho_{\mathrm{xc}}$ is given by%
\begin{equation}
\rho_{\mathrm{xc}}(r,r^{\prime},r^{\prime\prime};t)=[g(r,r^{\prime}%
,r^{\prime\prime};t)-1]\rho(r^{\prime\prime}),
\end{equation}
in which the correlator $g$ is defined according to%
\begin{align} 
&\left\langle T[\hat{\rho}(r^{\prime\prime}t)\hat{\psi}(rt)\hat{\psi
}^{\dag}(r^{\prime})]\right\rangle 
\nonumber\\
&=iG(r,r^{\prime};t)g(r,r^{\prime}%
,r^{\prime\prime};t)\rho(r^{\prime\prime})
\nonumber\\
&=iG(r,r^{\prime};t)[\rho(r'')+\rho_{\mathrm{xc}}(r,r^{\prime},r^{\prime\prime};t)]. \label{gh}%
\end{align}
The exchange-correlation hole fulfills the sum rule
\begin{equation}
\int d^{3}r^{\prime\prime}\rho_{\mathrm{xc}}(r,r^{\prime},r^{\prime\prime
};t)=-\delta_{\sigma\sigma^{\prime\prime}}\theta(-t) \label{sum-rule0}%
\end{equation}
and it has the property%
\begin{equation}
\rho_{\mathrm{xc}}(r,r^{\prime},r;t)=-\rho(r) \label{rho0}%
\end{equation}
for any $r,r^{\prime}$, and $t$.
The sum-rule and the above property may be seen as the dynamic version of the corresponding
properties of the static exchange-correlation
hole \cite{becke2014} originating from the seminal work of Slater on 
the exchange hole \cite{slater1951,slater1968}.
For a given $r'$, $V_{\mathrm{xc}}(r,r^{\prime};t)$ may be interpreted as 
a local time-dependent one-particle potential in which the added hole/electron moves.

In the language of functional derivative technique, 
the exchange-correlation hole can be related to the functional derivative of the Green function 
with respect to a probing field $\varphi$ \cite{hedin1965,aryasetiawan1998}. Since
\begin{align}
    \left\langle T[\hat{\rho}(3)\hat{\psi}(1)\hat{\psi
}^{\dag}(2)]\right\rangle &=i\rho(3)G(1,2) -\frac{\delta G(1,2)}{\delta \varphi(3)}
\end{align}
it follows that
\begin{equation}
    i\rho_\mathrm{xc} (1,2,3)G(1,2)=-\frac{\delta G(1,2)}{\delta \varphi(3)},
\end{equation}
where $1=(r_1,t_1)$ etc.

\subsection{General iterative solution}

From the equations of motion for $G$ and $G^\mathrm{H}$ of the Hartree approximation,
a Dyson-like equation can be constructed
as follows
\begin{align}
    &G(r,r';t) = G^\mathrm{H}(r,r';t) 
    \nonumber\\
    &+\int dr'' dt' G^\mathrm{H}(r,r'';t-t')
    V_{\mathrm{xc}}(r'',r';t') G(r'',r';t').
    \label{Dyson-like}
\end{align}
By operating $i\partial/\partial t - h(r)$ on both sides of the equation, it 
can be verified that the above $G$ fulfills the equation of motion.
This Dyson-like equation can be used as an iterative scheme for solving for $G$. The iteration
is started by setting $G=G^\mathrm{H}$ on the right-hand side and continued until 
self-consistency is achieved. One may choose a different starting point, 
such as the Kohn-Sham \cite{kohn1965} Green function
$G^{\mathrm{KS}}$, in which case $V_\mathrm{xc}$ must be replaced with 
$V_\mathrm{xc}-V^{\mathrm{KS}}_\mathrm{xc}$.

In practice, it is convenient to express the equation of motion in a set of
base orbitals $\{\varphi_{i}\}$:%
\begin{equation}
i\frac{\partial}{\partial t}G_{ij}(t)-\sum_{k}h_{ik}G_{kj}(t)-\sum
_{kl}V_{ik,lj}^{\mathrm{xc}}(t)G_{kl}(t)=\delta_{ij}\delta(t),
\end{equation}
where $G_{ij}$ and $h_{ik}$ are the matrix elements of $G$ and $h$ in the
orbitals and
\begin{equation}
V_{ik,lj}^{\mathrm{xc}}(t)=\int d^{3}rd^{3}r^{\prime}\;\varphi_{i}^{\ast
}(r)\varphi_{k}(r)V_{\mathrm{xc}}(r,r^{\prime};t)\varphi_{l}^{\ast}(r^{\prime
})\varphi_{j}(r^{\prime}).
\end{equation}
The Dyson-like equation becomes
\begin{equation}
    G_{ij}(t) = G^\mathrm{H}_{ij}(t) + \sum_{k} \int dt' G^\mathrm{H}_{ik}(t-t')F_{kj}(t'),
\end{equation}
in which
\begin{equation}
    F_{kj}(t') = \sum_{lm} V_{kl,mj}^{\mathrm{xc}}(t')G_{lm}(t').
\end{equation}
Fourier transformation leads to
\begin{equation}
    G_{ij}(\omega) = G^\mathrm{H}_{ij}(\omega)
    + \sum_{k}  G^\mathrm{H}_{ik}(\omega)F_{kj}(\omega),
\end{equation}
and
\begin{equation}
    F_{kj}(\omega) = \sum_{lm} \int \frac{d\omega'}{2\pi}
    V_{kl,mj}^{\mathrm{xc}}(\omega-\omega')G_{lm}(\omega').
\end{equation}
This yields an integral equation for $F(\omega)$:
\begin{align}
    F_{kj}(\omega) &= \sum_{lm} \int \frac{d\omega'}{2\pi}
    V_{kl,mj}^{\mathrm{xc}}(\omega-\omega')
    \nonumber\\
    &\times
    \left[ G^\mathrm{H}_{lm}(\omega') 
    +\sum_n G^\mathrm{H}_{ln}(\omega') F_{nm}(\omega') \right] .
    \label{Fomega}
\end{align}
Since $G^\mathrm{H}$ is a non-interacting Green function, the integrals over $\omega'$ can be
performed analytically.

Alternatively, by isolating the terms containing $G_{ij}$, the equation of motion can be
rewritten as follows:
\begin{equation}
\left\{  i\frac{\partial}{\partial t}-h_{ii}-V_{ii,jj}^{\mathrm{xc}}%
(t)-D_{ij}(t)\right\}  G_{ij}(t)=\delta_{ij}\delta(t),
\end{equation}
where 
\begin{align}
&D_{ij}(t)=\frac{1}{G_{ij}(t)}
\nonumber\\
&\times\left\{  \sum_{k\neq i}h_{ik}G_{kj}%
(t)+\sum_{k\neq i,l\neq j}V_{ik,lj}^{\mathrm{xc}}(t)G_{kl}(t)\right\}  .
\end{align}
The formal iterative solution is given by%
\begin{align}
&G_{ij}(t)= \left[G_{ij}(0^-)\theta(-t) +G_{ij}(0^+)\theta(t)\right]  
\nonumber\\
&\times\exp\left\{  -i\int_{0}^{t}dt^{\prime}\left[  h_{ii}%
+V_{ii,jj}^{\mathrm{xc}}(t^{\prime})+D_{ij}(t^{\prime})\right]  \right\},
\label{solution}%
\end{align}
where
\begin{equation}
    iG_{ij}(0^+) -iG_{ij}(0^-) = \delta_{ij},
\end{equation}
which is obtained by integrating the equation of motion from $0^-$ to $0^+$.
Since the Green
function to be solved appears on the right-hand side of the equation, the
formal solution in Eq. (\ref{solution}) facilitates an iterative scheme for
solving the Green function.

\subsection{Self-energy $vs.$ exchange-correlation potential}

A fundamental difference between the traditional self-energy and the
exchange-correlation potential approaches is that the former acts on the Green
function as a convolution in space and time whereas the latter acts
multiplicatively. The multiplicative property of $V_{\mathrm{xc}}$ has
consequences. 
One of these is that the equation of motion in Eq.
(\ref{EOM}) separates into the equation for the hole Green function $(t<0)$ and
for the electron Green function $(t>0)$.
In contrast, due to the convolution in time in the self-energy term,
solving for the hole Green function using the equation of motion in Eq.
(\ref{EOM-Sigma}) requires explicit knowledge of the electron Green function
and vice versa. 
On the other hand, the self-energy formalism is advantageous when expressed in
frequency space since the Dyson equation can be solved for each frequency whereas the
exchange-correlation potential formalism involves a convolution in frequency. 
Thus, the two approaches complement one another.

It was recently shown for the half-filled one-band Hubbard model in a square lattice that the use 
of the self-energy, calculated to a finite order of expansion in the interaction, 
to determine the Green function via the Dyson equation can lead to incorrect physics 
\cite{mcniven2021}. While a direct expansion of the Green function to the same order yields an 
insulating behavior, the Green function obtained from the Dyson equation results in a metallic 
behavior. This discrepancy can be traced back to the reducible diagrams implicitly summed when 
solving the Dyson equation, which generally differ from those in the direct expansion of 
the Green function at each order. This finding raises questions on the appropriateness of using 
the Dyson equation. The exchange-correlation potential formalism, on the other hand, is not meant 
to rely upon many-body perturbation theory but rather on direct construction based on known exact 
or accurate results of model systems and on exact properties of the exchange-correlation hole.

The choice of the self-energy as a truncation scheme may be understandable
from historical perspective. In the 1950's and early 1960's it was presumably
inconceivable to even consider many-body calculations on real materials. The
commonly used model of solids at that time was the electron gas. For the
electron gas, the definition of the self-energy acting on the Green function
as a convolution in space and time has a great advantage in that it allows for
a simple expression for the equation of motion or the Dyson equation when
Fourier transformed:%
\begin{equation}
G(k,\omega)=G_{0}(k,\omega)+G_{0}(k,\omega)\Sigma(k,\omega)G(k,\omega),
\end{equation}
where $G_0$ is the Green function of the free-electron gas.
The self-energy has since then been synonymous with the Green function and
become the accepted route for calculating the Green function to this day. 

The important role that the self-energy has played in
electronic structure theory is not to be undermined. 
In the self-energy formalism an iterative
equation for the self-energy can be derived yielding \cite{hedin1965,aryasetiawan1998}
\begin{align}
\Sigma(1,2) &  =i\int d3d4v(1-3)G(1,4)\nonumber\\
&  \times\left[  \left\{  \delta(4-3)+\frac{\delta V_\mathrm{H}(4)}{\delta\phi
(3)}\right\}  \delta(4-2)+\frac{\delta\Sigma(4,2)}{\delta\phi(3)}\right]  .
\end{align}
Here, $1=(r_{1}t_{1})$ etc. and $\phi$ is a probing field which is set to
zero after the derivative is taken. Neglecting the term $\delta\Sigma
/\delta\phi$ leads to the well-known $GW$ approximation. This kind of
iterative or perturbative equation is more difficult to establish for
$V_{\mathrm{xc}}$ since the equation of motion in Eq. (\ref{EOM}) cannot be
easily inverted to obtain $G^{-1}$.

From the definition of the self-energy in Eq. (\ref{def:Sigma}), it can be
seen that the Coulomb interaction has been lumped into the self-energy. While
it is formally and mathematically correct, the definition makes no use of the fact that the 
Coulomb interaction is known explicitly. The definition of the correlator $g$ in Eq. (\ref{gh}), 
on the other hand, is independent of the
Coulomb interaction. 
The special property of the Coulomb interaction being
dependent only on the distance between two electrons can be exploited leading to the conclusion 
that only the spherical average of the exchange-correlation hole is
relevant \cite{gunnarsson1976,jones1989}. 
This should greatly simplify the search for a good approximation for
the exchange-correlation hole or potential.

\section{Model systems}

To illustrate the form of the time-dependent exchange-correlation potential, the half-filled 
Hubbard dimer is considered. The exchange-correlation potential extracted from the Hubbard dimer
is then used as an approximate $V_{\mathrm{xc}}$ for the 1D Hubbard chain.

Another example is a simplified Holstein model,
describing a core electron coupled to a set of bosons
such as plasmons or phonons. This Hamiltonian is appropriate to
model solids in which the valence electrons are relatively delocalized, resembling electron gas.
The alkalis and $s$-$p$ semiconductors and insulators are examples of such systems.

The fourth example is the hydrogen atom. Although it is not a many-electron system it
illustrates explicitly the sum rule and condition fulfilled by the exchange-correlation hole.

\subsection{Hubbard dimer}

\begin{figure}[htp]
\centering
\includegraphics[trim=100 240 110 250, clip,width=0.9\columnwidth]{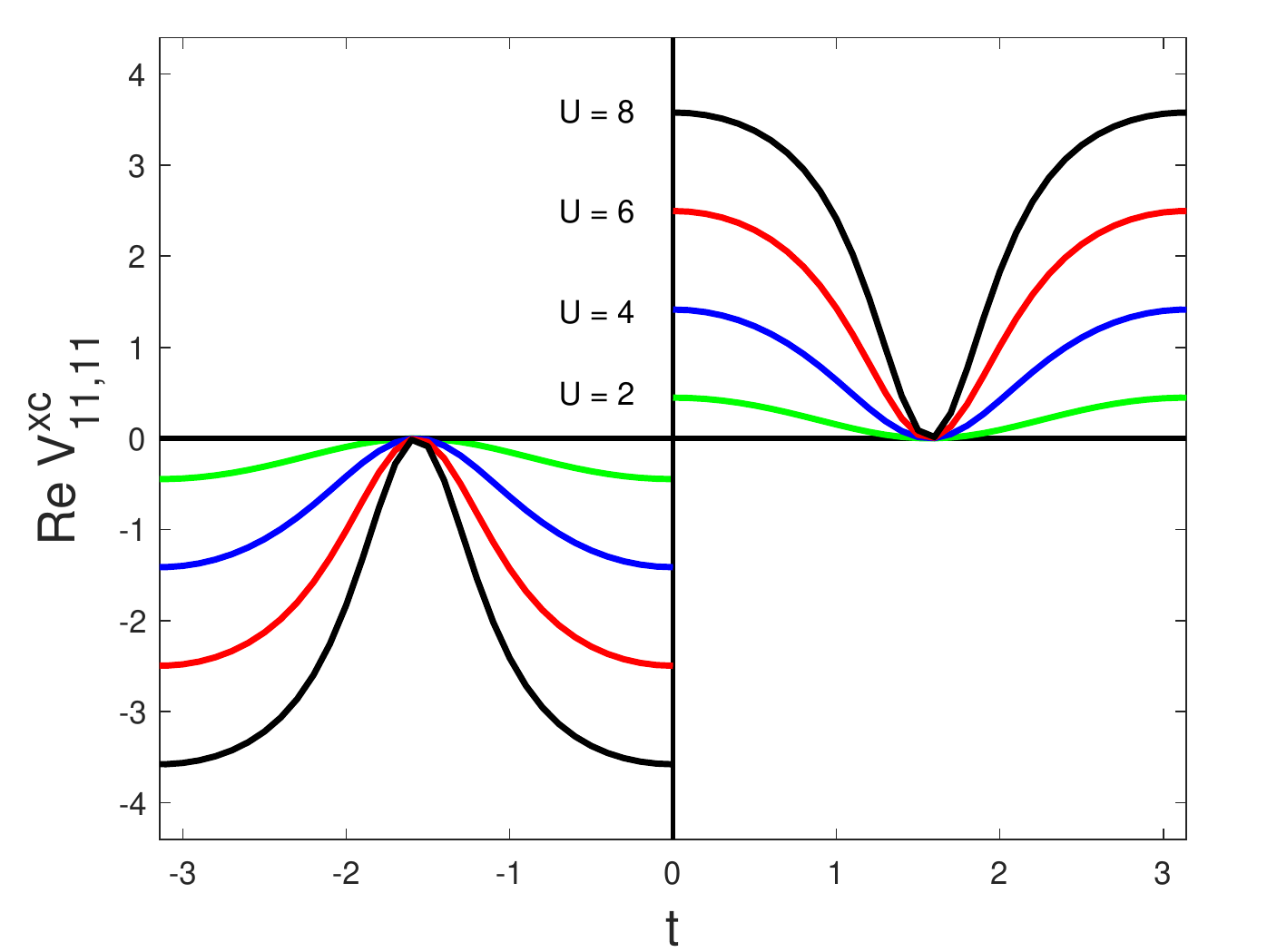}
\vfill
\hspace{1cm}
\includegraphics[trim=100 240 110 250, clip,width=0.9\columnwidth]{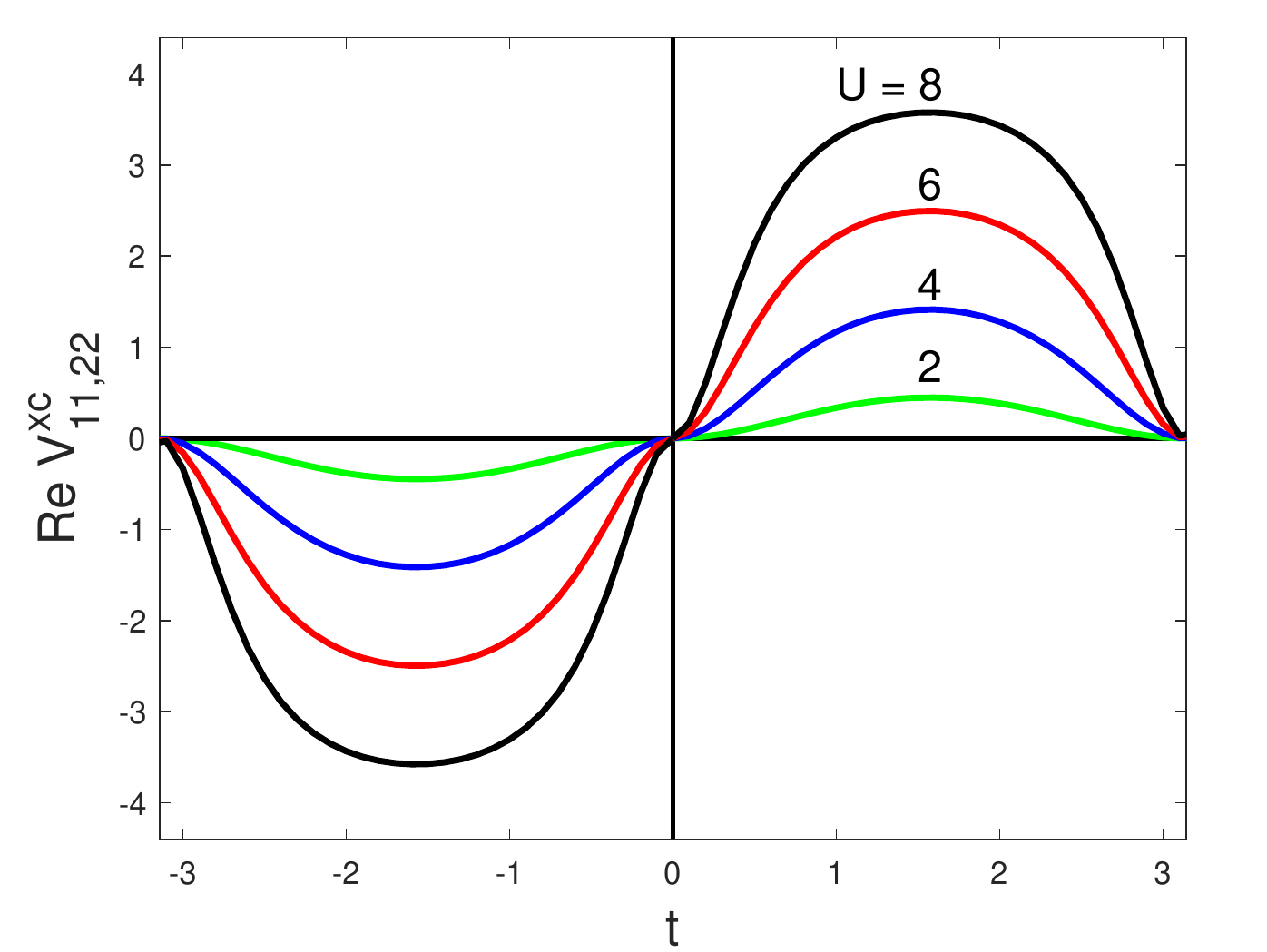}
\caption{The real parts of the exchange-correlation potentials 
$V^\mathrm{xc}_{11,11}$ and $V^\mathrm{xc}_{11,22}$ 
of the Hubbard dimer
as a function of time for $U=2,4,6,$ and $8$ with $\Delta=1$.
Due to the
particle-hole symmetry, $V_\mathrm{xc}(-t)=-V_\mathrm{xc}(t)$.}
\label{fig:ReVxc11}%
\end{figure}

\begin{figure}[htp]
\centering
\includegraphics[trim=100 240 110 250, clip,width=0.9\columnwidth]{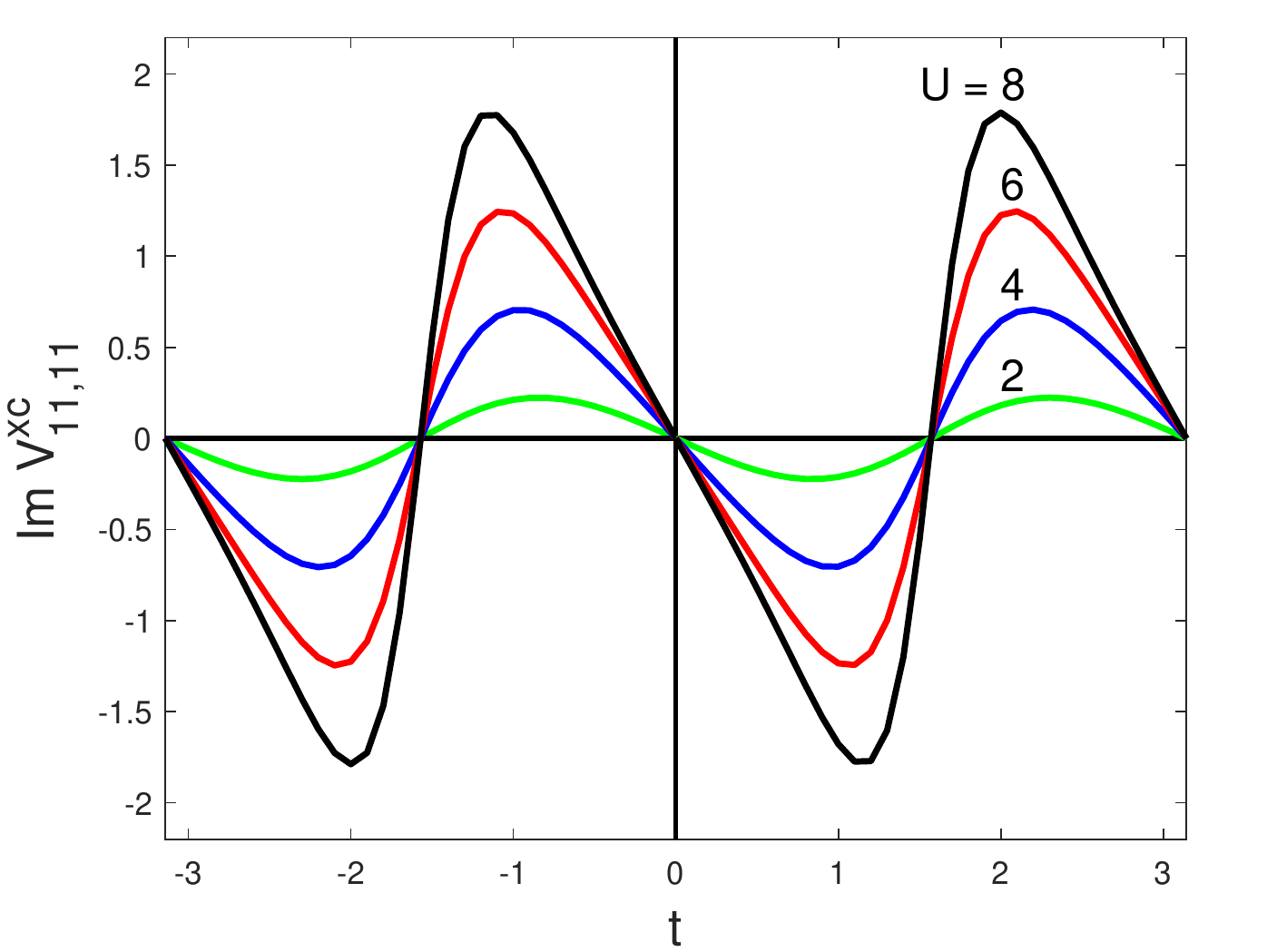}
\vfill
\hspace{1cm}
\includegraphics[trim=100 240 110 250, clip,width=0.9\columnwidth]{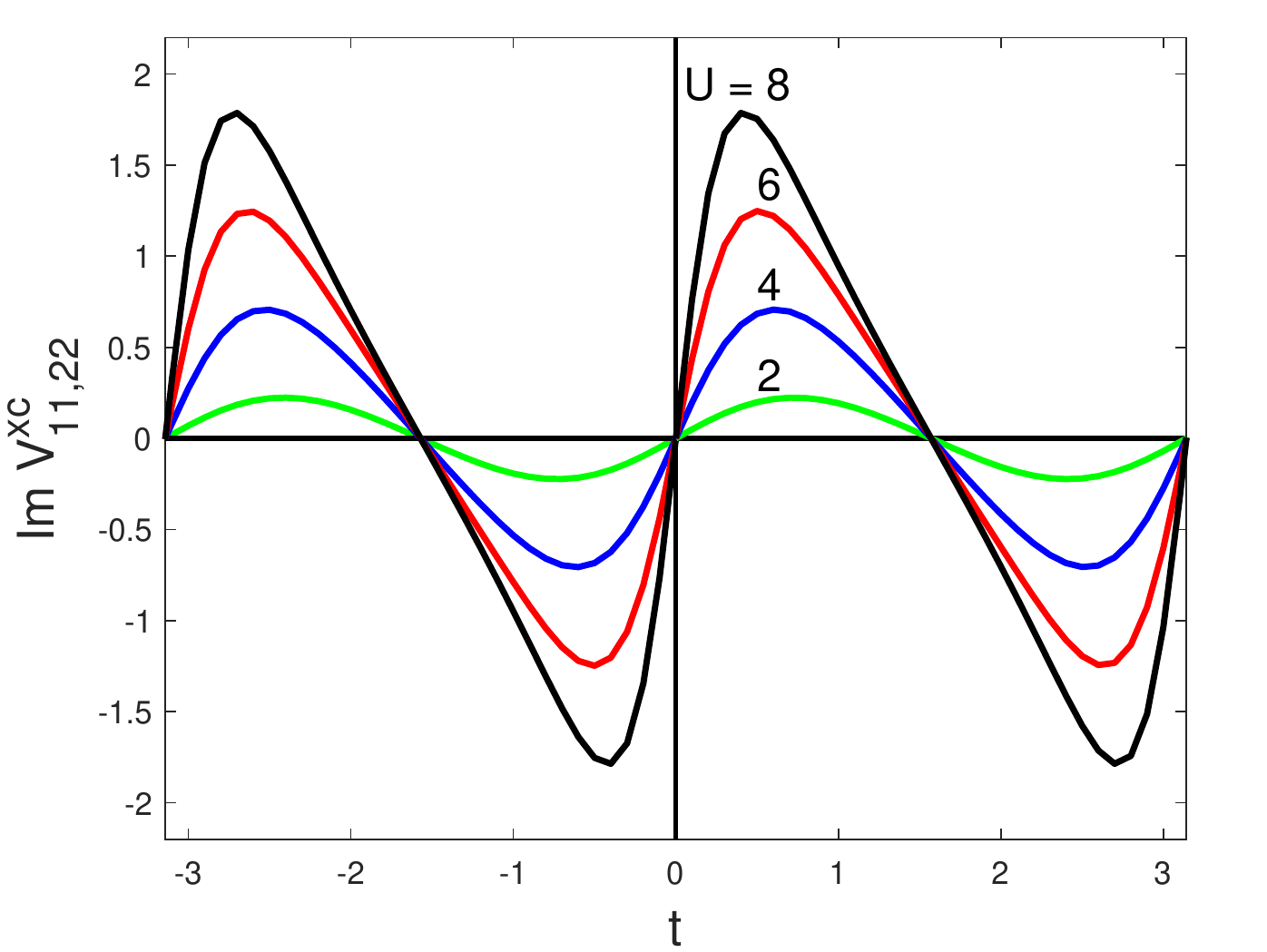}
\caption{The imaginary parts of the exchange-correlation potentials 
$V^\mathrm{xc}_{11,11}$ and $V^\mathrm{xc}_{11,22}$ 
of the Hubbard dimer
as a function of time for $U=2,4,6,$ and $8$ with $\Delta=1$.
Due to the
particle-hole symmetry, $V_\mathrm{xc}(-t)=-V_\mathrm{xc}(t)$.}
\label{fig:ImVxc11}%
\end{figure}

In the previous paper \cite{aryasetiawan2022}, the exchange-correlation potential of the 
half-filled Hubbard dimer was worked out analytically. 
The Hamiltonian of the Hubbard 
dimer in standard notation is given by%
\begin{equation}
\hat{H}=-\Delta\sum_{i\neq j}\hat{c}_{i\sigma}^{\dag}\hat{c}_{j\sigma}%
+U\sum_{i}\hat{n}_{i\uparrow}\hat{n}_{i\downarrow},\label{Hubbard}%
\end{equation}
where $i,j=1,2$. 
The results are given by
\begin{equation}
V_{11,11}^\mathrm{xc}(t>0)=\frac{\alpha U}{2}\frac{1+e^{-i2\Delta t}}{1+\alpha
^{2}e^{-i2\Delta t}}.
\label{Vxc1111}
\end{equation}%
\begin{equation}
V_{11,22}^\mathrm{xc}(t>0)=\frac{\alpha U}{2}\frac{1-e^{-i2\Delta t}}{1-\alpha
^{2}e^{-i2\Delta t}},
\label{Vxc1122}
\end{equation}
where%
\begin{equation}
\alpha=\frac{1-x}{1+x},
\label{alpha}
\end{equation}%
\begin{equation}
x=\frac{1}{4\Delta}\left(  \sqrt{U^{2}+16\Delta^{2}}-U\right)
\end{equation}
is the relative weight of double-occupancy configurations in the ground state
and%
\begin{equation}
2\Delta=E_{1}^{-}-E_{0}^{-}=E_{1}^{+}-E_{0}^{+}>0,
\end{equation}
are the excitation energies of the $(N\pm1)$-systems. From symmetry,
\begin{equation}
    V_{22,22}^\mathrm{xc}=V_{11,11}^\mathrm{xc},\;\;V_{22,11}^\mathrm{xc}=V_{11,22}^\mathrm{xc},
\end{equation}
\begin{equation}
  V_\mathrm{xc}(-t)=-V_\mathrm{xc}(t).  
\end{equation} 
For
convenience and for comparison, the results expressed in the site orbitals
shown in the previous article \cite{aryasetiawan2022} are shown in Figs. \ref{fig:ReVxc11} and 
\ref{fig:ImVxc11} but with different values of $U$.

It is perhaps more insightful to express
$V_{\mathrm{xc}}$ in the bonding and anti-bonding orbitals,
$V^\mathrm{xc}_\mathrm{AA,AA}$ and $V^\mathrm{xc}_\mathrm{BB,BB}$. In these orbitals,
the Green function is diagonal and $V_{\mathrm{xc}}$ in the bonding state is
identical to that in the anti-bonding one:
\begin{align}
    V^\mathrm{xc}(t>0)=
    &=\frac{1}{2}\left( V^\mathrm{xc}_{11,11}+V^\mathrm{xc}_{11,22}\right)
    \nonumber\\
    &=\frac{\alpha U}{2} 
    \frac{1-\alpha^2 e^{-i4\Delta t}}{1-\alpha^4 e^{-i4\Delta t}}.
    \label{VxcAA}
\end{align}
The other matrix element, $V^\mathrm{xc}_\mathrm{AB,BA}=V^\mathrm{xc}_\mathrm{BA,AB}$,
is given by
\begin{align}
    \Delta V^\mathrm{xc}(t>0)
    &=\frac{1}{2}\left( V^\mathrm{xc}_{11,11}-V^\mathrm{xc}_{11,22}\right)
    \nonumber\\
    &=\frac{\alpha U}{2} 
    \frac{(1-\alpha^2) e^{-i2\Delta t}}{1-\alpha^4 e^{-i4\Delta t}}.
    \label{VxcAB}
\end{align}

\begin{figure}[htp]
\centering
\includegraphics[trim=100 240 110 250, clip,width=0.9\columnwidth]{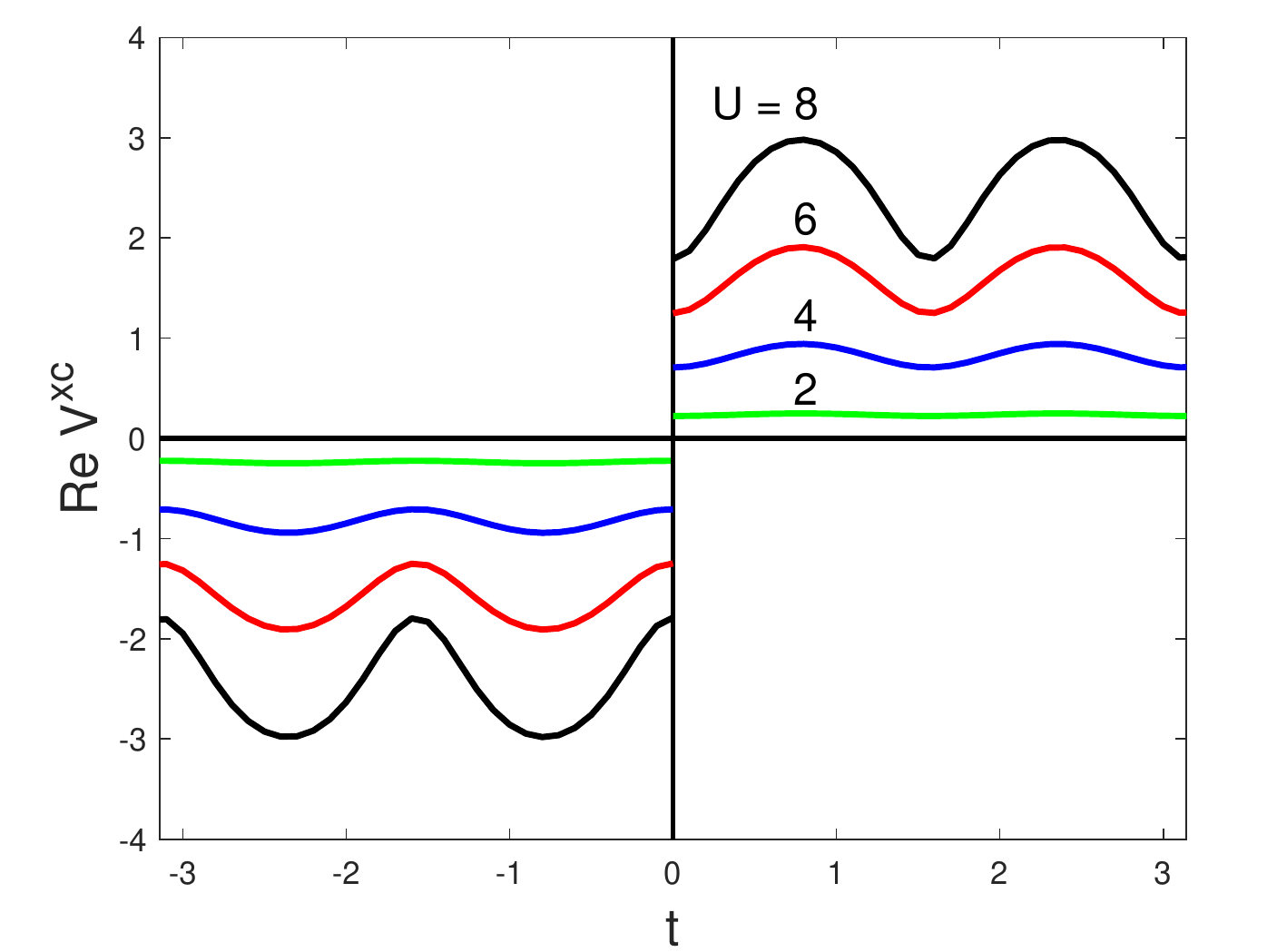}
\vfill
\hspace{1cm}
\includegraphics[trim=100 240 110 250, clip,width=0.9\columnwidth]{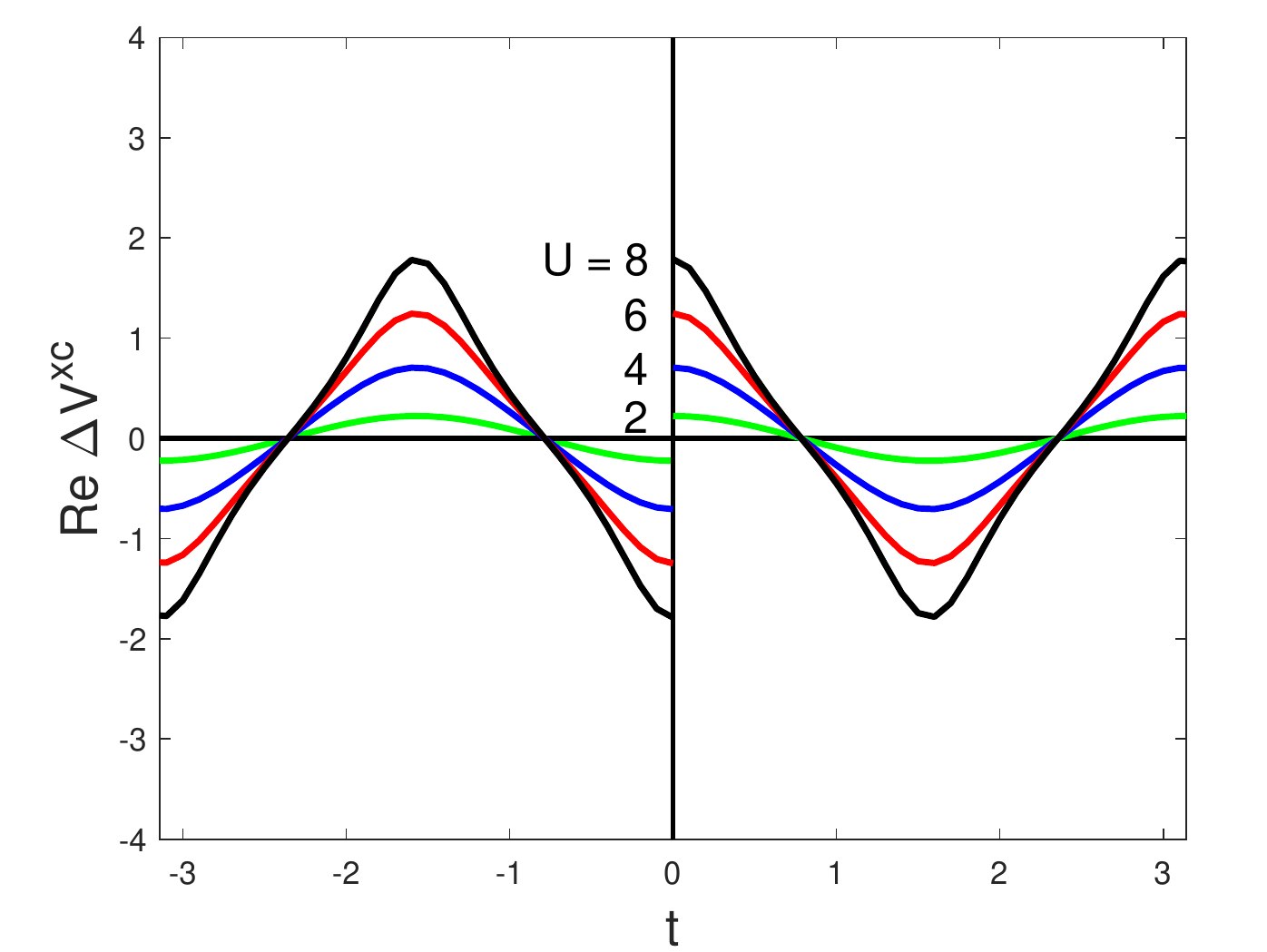}
\caption{The real parts of the exchange-correlation potential in
the bonding state $\phi_\mathrm{B}=\frac{1}{\sqrt{2}}(\varphi_1+\varphi_2)$
and anti-bonding state $\phi_\mathrm{A}=\frac{1}{\sqrt{2}}(\varphi_1-\varphi_2)$ as a function of 
time for $U=2,4,6,$ and $8$ with $\Delta=1$. 
}
\label{fig:ReVxc}%
\end{figure}

\begin{figure}[htp]
\centering
\includegraphics[trim=100 240 110 250, clip,width=0.9\columnwidth]{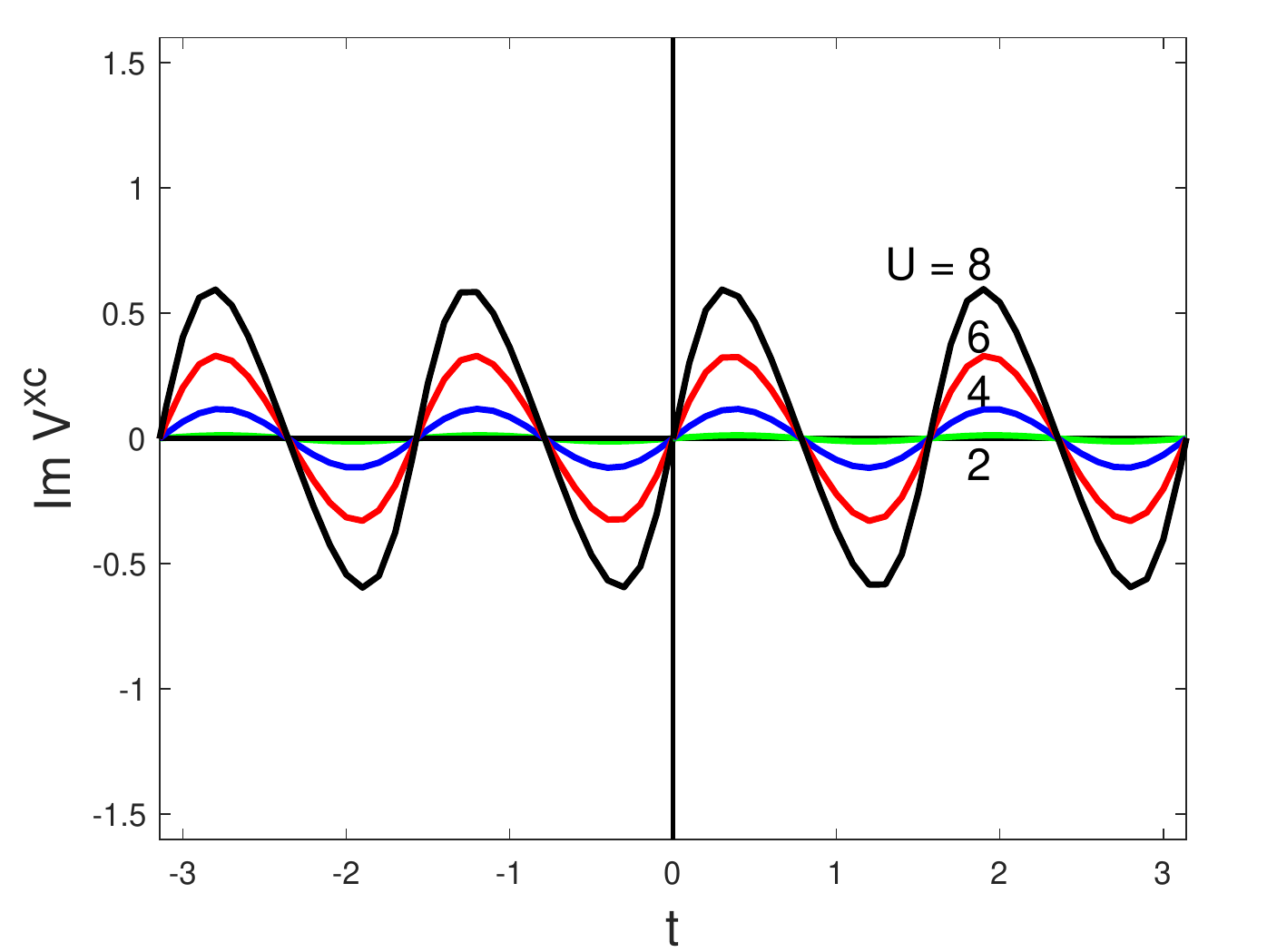}
\vfill
\hspace{1cm}
\includegraphics[trim=100 240 110 250, clip,width=0.9\columnwidth]{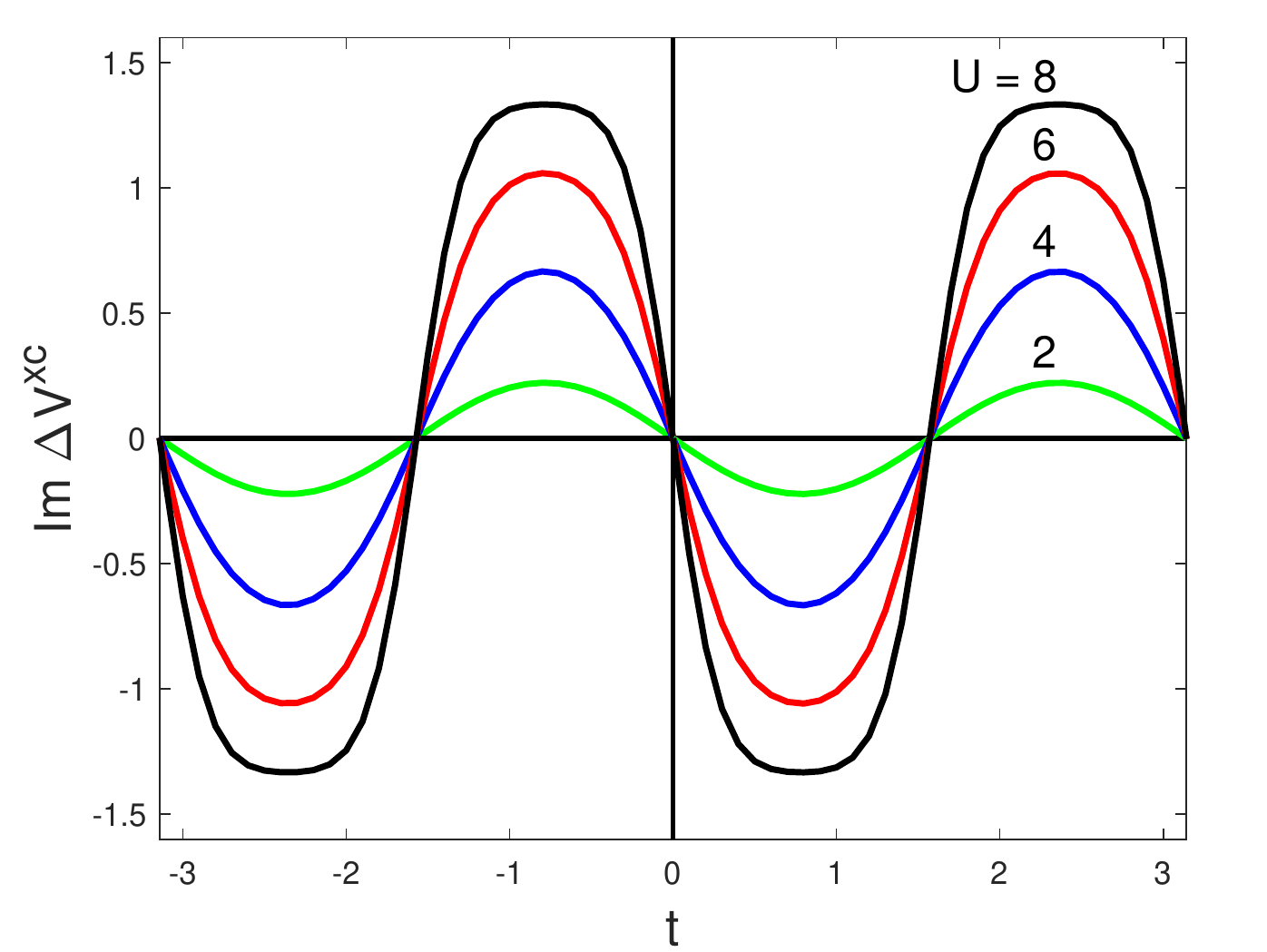}
\caption{The imaginary parts of the exchange-correlation potential in
the bonding and anti-bonding states as a function of time for $U=2,4,6,$ and $8$ with $\Delta=1$. 
}
\label{fig:ImVxc}%
\end{figure}

The results are shown in Fig. \ref{fig:ReVxc} for the real parts 
and in Fig. \ref{fig:ImVxc} for the imaginary parts.
The dependence of the correlation strength on time is revealed clearly, the stronger
$U$ the more pronounced the variation of $V_{\mathrm{xc}}$ with time. 

An interesting feature is the discontinuity of 
$V^\mathrm{xc}$ at 
$t=0$, which is the difference between the particle ($t=0^+$) and the hole ($t=0^-$)
values, reminiscent of the discontinuity in 
the exchange-correlation potential in density functional theory \cite{perdew1982}.

One also notices that the time dependence is dictated by the excitation
energies of the $(N\pm1)$ systems and in general, these excitations include
collective ones. For example, for solids one expects a time-dependent term of
the form $\exp(-i\omega_{p}t)$ where $\omega_{p}$ is the plasmon energy.
$V_\mathrm{xc}$ acts then as an effective external field, exchanging
an energy
$\omega_{p}$ with the system, as illustrated more explicitly later
in the example on the Holstein Hamiltonian. 

\subsection{1D Hubbard chain}

To calculate the Green function for the 1D Hubbard chain with the same
Hamiltonian as in Eq. (\ref{Hubbard}) but with $i,j=1,\dots ,\infty$, the
$V_{\mathrm{xc}}$ of the Hubbard dimer will be used as a model. This
approximation neglects components of $V_{\mathrm{xc}}$ beyond nearest neighbors.
Due to translational lattice symmetry it is convenient to introduce Bloch base
functions,%
\begin{equation}
\varphi_{k}(r)=\frac{1}{\sqrt{N}}\sum_{T}e^{ikT}\varphi(r-T),
\end{equation}
where $T$ denotes a lattice site.
The Green function expressed in these Bloch functions takes the form%
\begin{equation}
G(r,r^{\prime};t)=\sum_{k}\varphi_{k}(r)G(k,t)\varphi_{k}^{\ast}(r^{\prime
}),\label{Gkt}%
\end{equation}
where%
\begin{equation}
G(k,t)=\int drdr^{\prime}\varphi_{k}^{\ast}(r)G(r,r^{\prime};t)\varphi
_{k}(r^{\prime}).
\end{equation}

The equation of motion in the Bloch base functions is given by
\begin{equation}
    (i\partial_t -\varepsilon_q)G(q,t) -F(q,t) = \delta(t),
    \label{EOMqt}
\end{equation}
where
\begin{align}
    \varepsilon_q &= \int dr \varphi^*_q(r) h(r) \varphi_q(r)
    \nonumber\\
    &=\frac{1}{N}\sum_{TT'} e^{-ik(T-T')} \varphi^*(r-T) h(r) \varphi(r-T')
    \nonumber\\
    &=-2\Delta \cos{q}
\end{align}
and
\begin{align}
    &F(q,t) = \sum_k \int drdr' 
    \nonumber\\
    &\times \varphi_q^*(r)\varphi_k(r)V_{\mathrm{xc}}(r,r';t)
    \varphi_k^*(r')\varphi_q(r')
    \times G(k,t).
\end{align}
Since the Hartree potential is a constant it can be absorbed into the chemical potential.


To solve the equation of motion, one makes use of the $V^\mathrm{xc}$ deduced for the Hubbard
dimer. The equations of motion of the Green function in the bonding and anti-bonding orbitals
are given by
\begin{equation}
    \left[i\partial_t - \varepsilon_\mathrm{A} - V^\mathrm{xc}(t)\right]
    G_\mathrm{A}(t) -\Delta V^\mathrm{xc}(t)G_\mathrm{B}(t)=\delta(t),
\end{equation}
\begin{equation}
    \left[i\partial_t - \varepsilon_\mathrm{B} - V^\mathrm{xc}(t)\right]
    G_\mathrm{B}(t) -\Delta V^\mathrm{xc}(t)G_\mathrm{A}(t)=\delta(t),
\end{equation}
where
\begin{equation}
    \varepsilon_\mathrm{A}=\Delta,\;\;\;\varepsilon_\mathrm{B}=-\Delta.
\end{equation}
are the one-particle anti-bonding and bonding energies.
$V^\mathrm{xc}$ and $\Delta V^\mathrm{xc}$ are given, respectively, in Eqs.
(\ref{VxcAA}) and (\ref{VxcAB}).

With respect to the Hubbard chain, the bonding and anti-bonding states correspond to
the centers of the occupied and unoccupied bands. A physically motivated approximation
is to replace $G_\mathrm{A}(G_\mathrm{B})$ 
by $G(q,t)$ and $G_\mathrm{B}(G_\mathrm{A})$ by the average
$\frac{1}{N}\sum_k G(k,t)$:
\begin{equation}
    \left[i\partial_t - \varepsilon_q - V^\mathrm{xc}(t)\right]
    G(q,t) -\frac{1}{N}\sum_k \Delta V^\mathrm{xc}(t)G(k,t)=\delta(t).
\end{equation}
One may rewrite the above equation as follows:
\begin{equation}
    \left[i\partial_t - \varepsilon_q - V^\mathrm{xc}(t)
    -\Delta {V}^\mathrm{xc} (q,t)\right]
    G(q,t) =\delta(t),
\end{equation}
where 
\begin{equation}
    \Delta {V}^\mathrm{xc} (q,t) = \frac{1}{N}\sum_k \Delta V^\mathrm{xc}(t)
    \frac{G(k,t)}{G(q,t)}. 
    \label{DVxcApprox}
\end{equation}
The solution for the electron case is assumed to be given by 
\begin{equation}
    G^e(q,t)=-i\theta(t)e^{-i \varepsilon_q t - i\int_0^t dt'\left[V^\mathrm{xc}(t')
    +\Delta V^\mathrm{xc} (q,t') \right]}.
\end{equation}
A similar result can be readily derived for the hole Green function, keeping in mind that 
$V^\mathrm{xc}(-t)=-V^\mathrm{xc}(t)$.

To proceed further, the following approximation is proposed:
\begin{equation}
    \Delta V^\mathrm{xc} (q,t) \approx \frac{1}{N}\sum_k \Delta V^\mathrm{xc}(t)
    e^{-i(\varepsilon_k-\varepsilon_q)t},
\end{equation}
which corresponds to replacing $G$ by the non-interacting $G^0$.
Furthermore, to facilitate analytical calculations $V^\mathrm{xc}$ and $\Delta V^\mathrm{xc}$ 
in Eqs. (\ref{VxcAA}) and (\ref{VxcAB}) are
expanded as follows: 
\begin{equation}
    V^\mathrm{xc}(t) \approx \frac{\alpha U}{2},
\end{equation}
\begin{equation}
    \Delta V^\mathrm{xc}(t) \approx \frac{\alpha U}{2} (1-\alpha^2)
     e^{-i2\Delta t}.
\end{equation}
In the above approximation the constant term that shifts the one-particle energy
and the main excitation term $\exp{(-i2\Delta t)}$ that generates
the main satellites are kept whereas the higher excitation 
term $\exp{(-i4\Delta t)}$ is neglected.
One obtains to first order in $\Delta V^\mathrm{xc}$
\begin{align}
   & G^e(q,t) = -i\theta(t) e^{-i(\varepsilon_q+\frac{\alpha U}{2}) t}
    \nonumber\\
    &\times\left\{1  -i\frac{\alpha U}{2}(1-\alpha^2)\frac{1}{N}\sum_k \int_0^t dt' 
    e^{-i(\varepsilon_k-\varepsilon_q+2\Delta)t'}
     \right\}.
\end{align}

Performing the time integral one finds
\begin{align}
    &G^e(q,t) = -i\theta(t) e^{-i(\varepsilon_q+\frac{\alpha U}{2}) t}
    \nonumber\\
    &\times\left\{1+\frac{\alpha U}{2}(1-\alpha^2)\frac{1}{N}\sum_k 
    \frac{e^{-i(\varepsilon_k-\varepsilon_q+2\Delta)t }-1}{\varepsilon_k-\varepsilon_q+2\Delta} 
     \right\}.
\end{align}
Fourier transformation to the frequency domain yields
\begin{align}
    G^e(q,\omega) &= 
    \frac{A^e_0}{\omega-\left(\varepsilon_q+\frac{\alpha U}{2}\right)+i\eta}
    \nonumber\\
    &+\frac{1}{N}\sum_k \frac{A^e(k)}{\omega-(\varepsilon_k+\frac{\alpha U}{2}+2\Delta)+i\eta},
    \label{Geqt}
\end{align}
where
\begin{equation}
    A^e_0=1-\frac{\alpha U}{2}(1-\alpha^2)
    \frac{1}{N}\sum_k \frac{1}{\varepsilon_k-\varepsilon_q+2\Delta},
\end{equation}
\begin{equation}
    A^e(k)=\frac{\alpha U}{2} (1-\alpha^2)\frac{1}{\varepsilon_k-\varepsilon_q+2\Delta}.
\end{equation}

A similar derivation can be carried out for the hole Green function and the result is given by
\begin{align}
    &G^h(q,t) = i\theta(-t) e^{-i(\varepsilon_q-\frac{\alpha U}{2}) t}
    \nonumber\\
    &\times\left\{1-\frac{\alpha U}{2}(1-\alpha^2)\frac{1}{N}\sum_k 
    \frac{e^{-i(\varepsilon_k-\varepsilon_q-2\Delta)t }-1}{\varepsilon_k-\varepsilon_q-2\Delta} 
     \right\},
     \label{Ghqt}
\end{align}
\begin{align}
    G^h(q,\omega)& = 
    \frac{A^h_0}{\omega-\left(\varepsilon_q-\frac{\alpha U}{2}\right)-i\eta}
    \nonumber\\
    &-\frac{1}{N}\sum_k \frac{A^h(k)}{\omega-(\varepsilon_k-\frac{\alpha U}{2}-2\Delta)-i\eta},
    \label{Ghqw}
\end{align}
where
\begin{equation}
    A^h_0=1+\frac{\alpha U}{2}(1-\alpha^2)
    \frac{1}{N}\sum_k \frac{1}{\varepsilon_k-\varepsilon_q-2\Delta},
    \label{Ah0}
\end{equation}
\begin{equation}
    A^h(k)=\frac{\alpha U}{2}(1-\alpha^2) \frac{1}{\varepsilon_k-\varepsilon_q-2\Delta}.
    \label{Ahk}
\end{equation}
For the electron/hole case it is understood that both $\varepsilon_k$ and
$\varepsilon_q$ correspond to unoccupied/occupied states.

\begin{figure}[htp]
\centering
\includegraphics[trim=110 250 100 250, clip,width=\columnwidth]{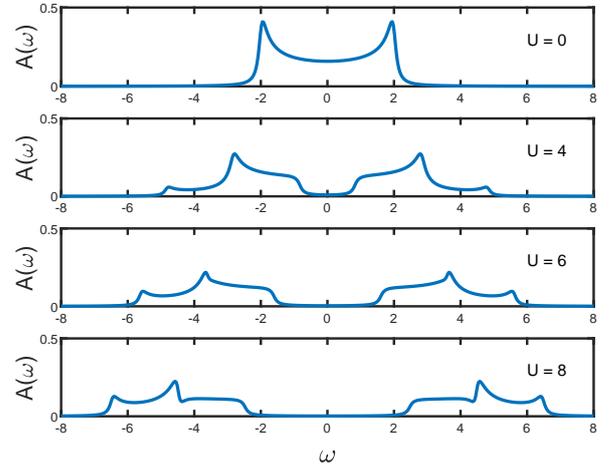}
\caption{The total spectral functions of the 1D Hubbard chain for $U=0,4,6,$ and $8$.
A broadening of $0.1$ has been used.}
\label{fig:Aw}%
\end{figure}

\begin{figure}[htp]
\centering
\includegraphics[trim=110 240 100 250, clip,width=\columnwidth]{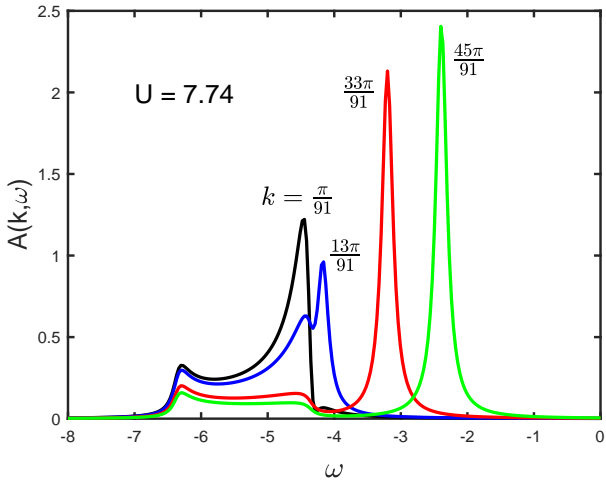}
\caption{The $k$-resolved spectral functions of the 1D Hubbard chain for $U=7.74$.
The $U$ and $k$ values have been chosen to facilitate comparison with Fig. 11 of 
Ref. \cite{benthien2007}. Within the approximation used, there is no spectral weight for 
$k=\frac{75\pi}{91}$ and $\frac{90\pi}{91}$ below the chemical potential, 
as explained in the text.
A broadening of $0.1$ has been used.}
\label{fig:AkwU7-74}%
\end{figure}

\begin{figure}[htp]
\centering
\includegraphics[trim=110 240 100 250, clip,width=\columnwidth]{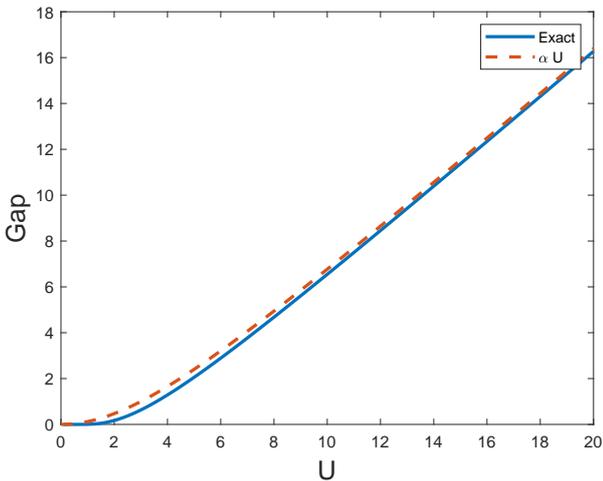}
\caption{The calculated band gap, $\alpha U$, 
compared with the exact gap obtained from the Bethe ansatz 
as a function of $U$. The calculated gap approaches the exact result as $U$ increases.}
\label{fig:Gap}%
\end{figure}

The calculated total spectral functions for $U = 2,4,6,$ and  $8$ are shown in Fig. \ref{fig:Aw}
and the $k$-resolved spectra are illustrated in Fig. \ref{fig:AkwU7-74} 
for a few $k$-points corresponding to $U=7.74$. This value has been chosen in order to make 
comparison with the results obtained using the dynamical density-matrix renormalization group 
method \cite{benthien2007}. Due to the approximation made in replacing
$G$ with $G^0$ in Eq. (\ref{DVxcApprox}), there is no spectral weight below the chemical
potential arising from the electron Green function and similarly, there is no spectral weight
above the chemical potential arising from the hole Green function. For this reason, 
there is no spectral weight below the chemical potential for 
$k=\frac{75\pi}{91}$ and $k=\frac{90\pi}{91}$
when compared with the results shown in Fig. 11 of Benthien and Jeckelmann \cite{benthien2007}.
Furthermore, the peaks are
sharp $\delta$-functions but with some weight transferred to higher or lower energy. It is known that for a one-dimensional interacting system, a removal/addition of an electron results
in two branches consisting of a spinless holon/antiholon dispersion
with a hole/electron charge and a charge-neutral spinon dispersion of spin $\frac{1}{2}$.

Although the approximation used is very simple, 
the $k$-resolved spectra shown in Fig. \ref{fig:AkwU7-74}
is in favorable agreement with those calculated using the dynamical 
density-matrix renormalization group 
method \cite{benthien2007}. A notable discrepancy is the dispersion widths for both the spinon
and the holon branches. Thus, the peak positions at small $k$ are lower in energy compared with
those in Ref. \cite{benthien2007}. 
This is understandable since within the approximation used, the dispersion
is pinned down by the non-interacting $G^0$ and the use of the Hubbard dimer $V^\mathrm{xc}$
neglects long-range correlations, which tend to narrow the band
dispersion.

The structure of the calculated $k$-resolved
hole spectra in Fig. \ref{fig:AkwU7-74} can be understood from 
the explicit expression for the hole Green function in Eq. (\ref{Ghqw}). The imaginary part of 
the first term is the main peak centered at $\omega=\varepsilon_q-\frac{\alpha U}{2}$, which is
interpreted as the spinon excitation. The imaginary part of the
second term with
the weight $A^h(k)$ set to unity is the non-interacting occupied
density of states shifted by $-(\frac{\alpha U}{2}+2\Delta)$.
For $q=\frac{45\pi}{91}$ (green curve),
the one-particle energy, $\varepsilon_q$, is approximately zero and the main peak is located
at the top of the band edge at $\omega=-\frac{\alpha U}{2}$. The remaining spectra stretching
approximately from $-6.5$ to $-4.5$ is the shifted occupied 
density of states weighted by $A^h(k)$.
The peak feature at the lower edge at $-6.5$ reflects the same feature present in
the shifted 
non-interacting occupied density of states ($U=0$ in Fig. \ref{fig:Aw})
and may be interpreted as the holon 
excitation. The structure at the upper edge at around $-4.5$, 
on the other hand, arises from the enhanced
weight of $A^h(k)$ which is largest for $q$ around
the bottom of the non-interacting band, giving rise to a step-like structure \cite{benthien2007}.
For $q=\frac{\pi}{91}$ (black curve),
the main peak is centered at $\omega=-\frac{\alpha U}{2}-2\Delta$ and it merges with
the shifted density of states.  For $q=\frac{13\pi}{91}$ the main peak still lies outside
the shifted density of states and a three-peak structure is then observed, as also found in
the calculated spectra using the dynamical density-matrix renormalization 
group method \cite{benthien2007}. 
It is not immediately evident from the expressions for the Green function is Eqs. (\ref{Geqt})
and (\ref{Ghqt}) how to interpret the two branches as 
spinon and holon excitations. Such interpretation would probably require analysis in terms of
state vectors.

The calculated band gap, which is given by 
$\alpha U$, is displayed in Fig. \ref{fig:Gap} and compared with the exact result 
\cite{ovchinnikov1970},
\begin{equation}
    E_\mathrm{gap} = \frac{16\Delta^2}{U} \int_1^\infty dy
    \frac{\sqrt{y^2-1}}{\sinh{\left(\frac{2\pi\Delta y}{U}\right)} },
    \label{ExactGap}
\end{equation}
obtained from the Bethe ansatz. The agreement between the analytically calculated gap
and the exact gap is very close, despite the simplicity of the approximation. 
The calculated gap approaches the exact gap in the limit of large $U$ and
a gap opens as soon as $U$ is finite, as in the exact case. The discrepancy is largest
at small values of $U$, which may be understood from the lack of long-range
screening effects in 
$V^\mathrm{xc}$ of the Hubbard dimer. It should be noted 
that in the widely used dynamical mean-field theory (DMFT) \cite{georges1996} 
within the single-site approximation, the gap is not opened up until 
$U > 6$. Only within the cluster DMFT with an even number of sites does the gap form for 
any finite $U$ whereas with an odd number of sites a metallic region remains until $U$ exceeds
a certain value before entering the Mott insulating phase through a coexistence region
\cite{go2009}. 

\subsection{Holstein Hamiltonian}

A simplified Holstein Hamiltonian describing a coupling between a core electron and a set
of bosons, such as plasmons or phonons, is given by \cite{langreth1970}
\begin{equation}
    \hat{H} = \varepsilon \hat{c}^\dagger \hat{c}
    +\sum_q \hat{c} \hat{c}^\dagger g_q(\hat{b}_q + \hat{b}_q^\dagger)
    +\sum_q \omega_q \hat{b}_q^\dagger \hat{b}_q,
\end{equation}
where $\varepsilon$ is the core electron energy, $\omega_q$ 
is the boson energy of wave vector $q$,
and $\hat{c}$ and $\hat{b}_q$ are respectively the core electron and the boson operators. 
This Hamiltonian can be solved analytically and the algebra is simplified if it is assumed that
the boson is dispersionless with an average energy $\omega_p$. Under this assumption, the
exact solution for the core-electron removal spectra yields \cite{langreth1970}
\begin{equation}
    A(\omega)=\sum_{n=0}^\infty f_n \delta(\omega-\varepsilon-\Delta\varepsilon+n\omega_p),
\end{equation}
where
\begin{equation}
    f_n=\frac{e^{-a}a^n}{n!},\;\;a=\sum_q \left( \frac{g_q}{\omega_p}\right)^2, 
    \;\; \Delta\varepsilon = a\omega_p.
\end{equation}
This exact solution can also be obtained using the cumulant expansion
\cite{bergersen1973,hedin1980,almbladh1983}. 
The hole Green function corresponding to the above spectra is given by
\begin{equation}
    G(t<0) = i \sum_{n=0}^\infty f_n e^{-i(\varepsilon+\Delta\varepsilon-n\omega_p)t}\theta(-t).
\end{equation}
It can be verified that the time-dependent exchange-correlation potential corresponding to
the Holstein Hamiltonian reads
\begin{equation}
    V_\mathrm{xc}(t<0)= \Delta\varepsilon \left(1-e^{i\omega_p t}\right).
\end{equation}
This expression provides a very simple interpretation: the first term corrects the non-interacting
core-electron energy whereas the second term describes the bosonic mode interacting
with the core electron, which can exchange not only one but multiple quanta of $\omega_p$ with
the field.
This is precisely what is accomplished by the cumulant expansion
\cite{bergersen1973,hedin1980,almbladh1983}
within the self-energy formulation but in an \emph{ad hoc} and complicated manner.

\subsection{Hydrogen atom}

It may seem trivial to consider the hydrogen atom since it is not a
many-electron system. Nevertheless, it illustrates a number of exact results
such as the sum rule in Eq. (\ref{sum-rule0}) and the condition in Eq.
(\ref{rho0}). For the hydrogen atom, the hole Green function is given by%
\begin{equation}
G(r,r^{\prime};t<0)=i\varphi_{s}(r)\varphi_{s}(r^{\prime})\exp(-i\varepsilon
_{s}t)\theta(-t),
\end{equation}
where $\varphi_{s}$ and $\varepsilon_{s}$ are the $1s$-orbital and its energy.
The exchange-correlation potential for $t<0$ can be readily deduced yielding%
\begin{equation}
V_{\mathrm{xc}}(r,r^{\prime};t<0)=-V_{\mathrm{H}}(r)=-\int dr^{\prime\prime
}v(r-r^{\prime\prime})|\varphi_{s}(r^{\prime\prime})|^{2},
\end{equation}
independent of $r^{\prime}$ and $t$, cancelling the
spurious Hartree potential. The corresponding exchange-correlation hole is
then given by%
\begin{equation}
\rho_{\mathrm{xc}}(r,r^{\prime},r^{\prime\prime};t<0)=-|\varphi_{s}%
(r^{\prime\prime})|^{2}.
\end{equation}
It can also be seen that the condition in Eq. (\ref{rho0}),%
\begin{equation}
\rho_{\mathrm{xc}}(r,r^{\prime},r;t<0)=-\rho(r),
\end{equation}
as well as the sum rule in Eq. (\ref{sum-rule0}),%
\begin{equation}
\int dr^{\prime\prime}\rho_{\mathrm{xc}}(r,r^{\prime},r^{\prime\prime
};t<0)=-\int dr^{\prime\prime}|\varphi_{s}(r^{\prime\prime})|^{2}=-1,
\end{equation}
are both fulfilled, as they should be.

\section{Conclusion}

The exchange-correlation formalism has been applied to determine
the Green function of
the 1D Hubbard chain by utilizing the exchange-correlation potential derived from the Hubbard
dimer. Under the approximation corresponding to replacing the full Green function by a 
non-interacting Green function in the first iteration, the spectral functions can be calculated
analytically. Despite the very simple approximation, the results compare favorably with
the more accurate results
calculated using the dynamical density-matrix renormalization group method.
Peak structures corresponding to the holon and spinon collective excitations
are correctly reproduced, although the positions are too low for small $k$, due
to the use of a non-interacting Green function and the neglect of long-range correlations in
the exchange-correlation potential of the Hubbard dimer. 
The calculated gap agrees very well with the exact result obtained from
the Bethe ansatz. These very encouraging results may indicate the robustness of the 
exchange-correlation potential, insensitive to the system size, 
allowing for extrapolation from a small to a large system. By using the exact $V^\mathrm{xc}$
of the Hubbard dimer, it is ensured that the sum-rule is fulfilled.

An example from the Holstein Hamiltonian further illustrates the potential of the 
exchange-correlation formalism. The exact $V^\mathrm{xc}$ has a very simple form,
offering a clear physical interpretation. The main collective charge excitation (plasmon)
determines the characteristic energy of the time-dependent part of $V^\mathrm{xc}$ while the
constant term provides a correction to the one-particle energy. One may speculate that a generic 
structure of the exchange-correlation potential consists of a constant term and a series of
time-dependent terms of exponential form
with energies characteristic of the excitations of the $(N\pm 1)$-systems.

An example of the hydrogen atom illustrates the
exact properties of the exchange-correlation hole.

\begin{acknowledgments}
Financial support from the Knut and Alice Wallenberg  
Foundation (KAW 2017.0061) and the Swedish Research Council 
(Vetenskapsrådet, VR 2021-04498\_3) is gratefully acknowledged. 
\end{acknowledgments}


\end{document}